\documentclass[english,aps, prb, superscriptaddress, twocolumn, showpacs,nofootinbib, a4paper]{revtex4}
\usepackage[T1]{fontenc}

\usepackage[latin9]{inputenc}
\usepackage{color}
\usepackage{babel}
\usepackage{amsmath}
\usepackage{amssymb}
\usepackage{graphicx}
\usepackage{hyperref}

\makeatletter

\newcommand{\lyxdot}{.}
\renewcommand{\vec}[1]{\ensuremath{\mathbf{#1}}}
\newcommand{\zeroplus}{\ensuremath{{\tau\!=\!0^{+}}}}


\begin{document}

\title{
  Influence of Fock exchange in combined many-body perturbation and dynamical mean field theory
}
\author{Thomas Ayral}
\affiliation{Institut de Physique Théorique (IPhT), CEA, CNRS, UMR 3681, 91191 Gif-sur-Yvette, France}
\affiliation{Department of Physics and Astronomy, Rutgers University, Piscataway, NJ 08854, USA}
\author{Silke Biermann}
\affiliation{Centre de Physique Théorique, Ecole Polytechnique, CNRS, Université Paris Saclay, 91128 Palaiseau, France}
\author{Philipp Werner}
\affiliation{Department of Physics, University of Fribourg, 1700 Fribourg, Switzerland}
\author{Lewin Boehnke}
\affiliation{Department of Physics, University of Fribourg, 1700 Fribourg, Switzerland}

\begin{abstract}
In electronic systems with long-range Coulomb interaction, the nonlocal Fock exchange term has a band-widening effect. While this effect is included in combined many-body perturbation theory and dynamical mean field theory schemes, it is not taken into account in standard extended DMFT (EDMFT) calculations.
Here, we include this instantaneous term in both approaches and investigate its effect on the phase diagram and dynamically screened interaction. 
We show that the largest deviations between previously presented EDMFT and $GW$+EDMFT results originate from the nonlocal Fock term, and that the quantitative differences are especially large in the strong-coupling limit.
Furthermore, we show that the charge-ordering phase diagram obtained in $GW$+EDMFT methods for moderate interaction values is very similar to the one predicted by dual boson methods that include the fermion-boson or four-point vertex.
\end{abstract}

\maketitle

Dynamical Mean Field Theory\cite{Georges1996} (DMFT) 
self-consistently maps a correlated Hubbard lattice problem with local interactions onto an effective impurity problem 
consisting of a correlated orbital hybridized with a noninteracting 
fermionic bath. If the bath is integrated out, one obtains an impurity action with retarded hoppings. 
Extended dynamical mean field theory\cite{Sachdev1993, Sengupta1995,Kajueter1996,Si1996,Parcollet1999, Chitra2000,Chitra2001,Smith2000,Si2001}
(EDMFT)  
extends the DMFT idea to systems with long-range interactions. It does so by mapping a lattice problem with long-range interactions onto an effective impurity model with
self-consistently determined fermionic and bosonic baths, or, in the action formulation, an impurity model with retarded hoppings and retarded interactions.

While EDMFT captures dynamical screening effects and charge-order instabilities, it has  been found to suffer from qualitative shortcomings in finite dimensions. 
For example, the charge susceptibility computed in EDMFT does not coincide with the derivative of the average charge with respect to a small applied field\cite{vanLoon2015}, nor does it obey local charge conservation rules\cite{Hafermann2014} essential for an adequate description of collective modes such as plasmons.

The EDMFT formalism has an even more basic deficiency: since it is based on a local approximation to the self-energy, it does not include even the first-order nonlocal interaction term, the Fock term. The combined $GW$+EDMFT 
\cite{Biermann2003,Sun2004,Ayral2013} scheme corrects this by supplementing the local self-energy from EDMFT with the nonlocal part of the $GW$ diagram, where $G$ is the interacting Green's function and $W$ the fully screened interaction. %
Indeed, the nonlocal Fock term ``$[Gv]^\mathrm{nonloc}$'' is
included in the nonlocal ``$[GW]^\mathrm{nonloc}$'' diagram. As described in more detail in Ref.~\onlinecite{Ayral2013} (see also the appendix of Ref.~\onlinecite{Golez2017}), the $GW$+EDMFT method 
is formally obtained by constructing an energy functional of $G$ and $W$, the Almbladh\cite{Barth1999} functional $\Psi$, and by approximating $\Psi$ as a sum of two terms, one containing all local diagrams (corresponding to EDMFT), the other containing the simplest nonlocal correction (corresponding to the $GW$ approximation\cite{Hedin1965}). This functional construction rules out double-counting of local terms in the self-energy and polarization\cite{Biermann2004a, Ayral2013}.
Even though it has been introduced under the name $GW$+DMFT
\cite{Biermann2003} in the literature, we will denote this full scheme by $GW$+EDMFT
to emphasize that it is based on the EDMFT formalism, and to distinguish it 
from simplified implementations without two-particle self-consistency, 
which have appeared in the literature (and which we will denote in the
following as $GW$+DMFT).

In a recent implementation of the $GW$+EDMFT method, Ref.~\onlinecite{Ayral2013}, and related papers\cite{Ayral2012,Huang2014}, the nonlocal Fock term was omitted.\cite{Ayral2016a}  
Here, we explore and 
highlight the role of this term and its interplay with the local correlations. We quantify the band-widening effect of the Fock term and study the consequences of its presence or absence on various observables, and on the charge-order phase boundary. 
Our self-consistent implementation goes beyond previous studies of the effect of the Fock exchange
in realistic calculations, where it was studied systematically
within $GW$ \cite{Miyake2013} and $GW$+DMFT,\cite{Tomczak2014} 
albeit not in a self-consistent way.

The manuscript is organized as follows:
In section~\ref{sec:formalism}, we recap the $GW$+EDMFT equations with special emphasis on the Fock term and make general statements about the expected impact.
In section~\ref{sec:theeffectivebandstructure}, we show explicit results for the effective renormalization of the band structure by the instantaneous Fock contribution within $GW$+EDMFT, followed by systematic comparisons 
with the results of simplified 
formalisms in section~\ref{sec:withandwithoutfock}.
Section~\ref{sec:insidethemottphase} discusses the role of the Fock term in the Mott-insulating phase, where it stays relevant up to very large values of the on-site interaction.
Finally, in section~\ref{sec:DB}, we compare our results with 
results obtained within 
the recent dual boson method.

\section{Formalism}
\label{sec:formalism}
We aim at solving the extended Hubbard model on the two-dimensional square lattice 
by constructing an effective impurity problem that gives the local part of the self-energy~$\Sigma$ and polarization~$P$, and a diagrammatic expansion in their nonlocal components.
The model is defined by
the Hamiltonian
\begin{equation} \mathcal{H}=-\sum_{ij}t_{ij}c^\dagger_ic_j+\frac{1}{2}\sum_{ij} v_{ij}n_in_j-\mu\sum_in_i.
  \label{eq:eHubbard}
\end{equation}
Here,
$t_{ij}$ are the real-space hopping matrix elements, $c_i^{(\dagger)}$ the electronic annihilator (creator) on site $i$, $v_{ij}$ the Coulomb interaction and $n_i=c^\dagger_ic_i$.
We will
restrict ourselves to models with hoppings and interactions between nearest neighbors and next nearest neighbors only,
\begin{align}
  \label{eq:tUV}
  t_{ij}=&t\delta_{\left<ij\right>}+t'\delta_{\left<\left<ij\right>\right>},\\
  v_{ij}=&U\delta_{ij}+V\delta_{\left<ij\right>}+V'\delta_{\left<\left<ij\right>\right>},
\end{align}
where $\delta_{ij}$ is the usual Kronecker delta, $\delta_{\left<ij\right>}$(resp. $\delta_{\left<\left<ij\right>\right>}$) is 1 for $i$ and $j$ nearest neighbors (resp.  next-nearest neighbors, along the diagonal of the square lattice) and 0 otherwise.
This results in the Fourier transforms 
\begin{align}
  \varepsilon_{\vec{k}}=\phantom{+}&2t(\cos(k_x)+\cos(k_y))\nonumber\\
               &+2t'(\cos(k_x+k_y)+\cos(k_x-k_y))
  \label{eq:epsk}
\end{align}
and
\begin{align}
  v_{\vec{q}}=U&+2V(\cos(q_x)+\cos(q_y))\nonumber\\
               &+2V'(\cos(q_x+q_y)+\cos(q_x-q_y)).
  \label{eq:Vq}
\end{align}

The full expression for the self-energy in the $GW$+EDMFT approximation is
\begin{equation}
\Sigma(\vec{k},i\omega_n)=\Sigma_{\mathrm{imp}}(i\omega_n)+\Sigma_{GW_{c}}^{\mathrm{nonloc}}(\vec{k},i\omega_n)+\Sigma_{\mathrm{F}}^{\mathrm{nonloc}}(\vec{k}).\label{eq:Sigma_expr}
\end{equation}
The last two terms correspond to the nonlocal part of the $GW$ self-energy. They can be expressed as a function of imaginary time $\tau$ 
 and momentum $\vec{k}$ as follows:
\begin{eqnarray}
\Sigma_{GW_{c}}^{\mathrm{nonloc}}(\vec{k},\tau) & = & -\sum_{\vec{q}}G_{\vec{q+k}}(\tau)W^c_{\vec{q}}(\tau)\label{eq:Sigma_GWc}\\
 &  & +\left[\sum_{\vec{q}}G_{\vec{q+k}}(\tau)W_{\vec{q}}^{c}(\tau)\right]_{\mathrm{loc}},\nonumber \\
\Sigma_{\mathrm{F}}^{\mathrm{nonloc}}(\vec{k}) & = & -\sum_{\vec{q}}G_{\vec{q+k}}(\zeroplus)v_{\vec{q}}\label{eq:Sigma_Gv}\\
 &  & +\left[\sum_{\vec{q}}G_{\vec{q+k}}(\zeroplus)v_{\vec{q}}\right]_{\mathrm{loc}}.\nonumber 
\end{eqnarray}
Fourier transformations between $\tau$ and fermionic [bosonic] Matsubara frequencies $i\omega_n=i(2n+1)\frac{\pi}{\beta}$ [$i\nu_m=i2m\frac{\pi}{\beta}$] are assumed where needed ($\beta$ denotes the inverse temperature).
The ``loc'' suffix denotes a sum over the first Brillouin zone.
\begin{equation}
  \label{eq:G}
  G_{\vec{k}}(i\omega_n)=(i\omega_n+\mu-\varepsilon_\vec{k}-\Sigma(\vec{k},i\omega_n))^{-1}
\end{equation}
is the interacting lattice Green's function, and $W^{\mathrm{c}}$ is defined as
\begin{equation}
W_{\vec{q}}^{\mathrm{c}}(i\nu_m)\equiv\frac{v_{\vec{q}}}{1-v_{\vec{q}}P_{\vec{q}}(i\nu_m)}-v_{\vec{q}},
\end{equation}
with $P$ the polarization function.
All results are given in units of $D=4|t|$ (which is the half bandwidth when $t'=0$), and the momentum discretization is $N_{k}=32\times32$
points in the first Brillouin zone, 
unless otherwise stated. 
We use the original formulation of the $GW$+EDMFT scheme,\cite{Biermann2003} corresponding
-- within a functional formulation -- to a Hubbard-Stratonovich decoupling
of the full interaction term, dubbed ``HS-$UV$ decoupling'' in
Ref.~\onlinecite{Ayral2013}.
As argued there, this choice has the advantage that it treats local and 
nonlocal interactions on the same footing.

\begin{figure}
\includegraphics[width=1\columnwidth]{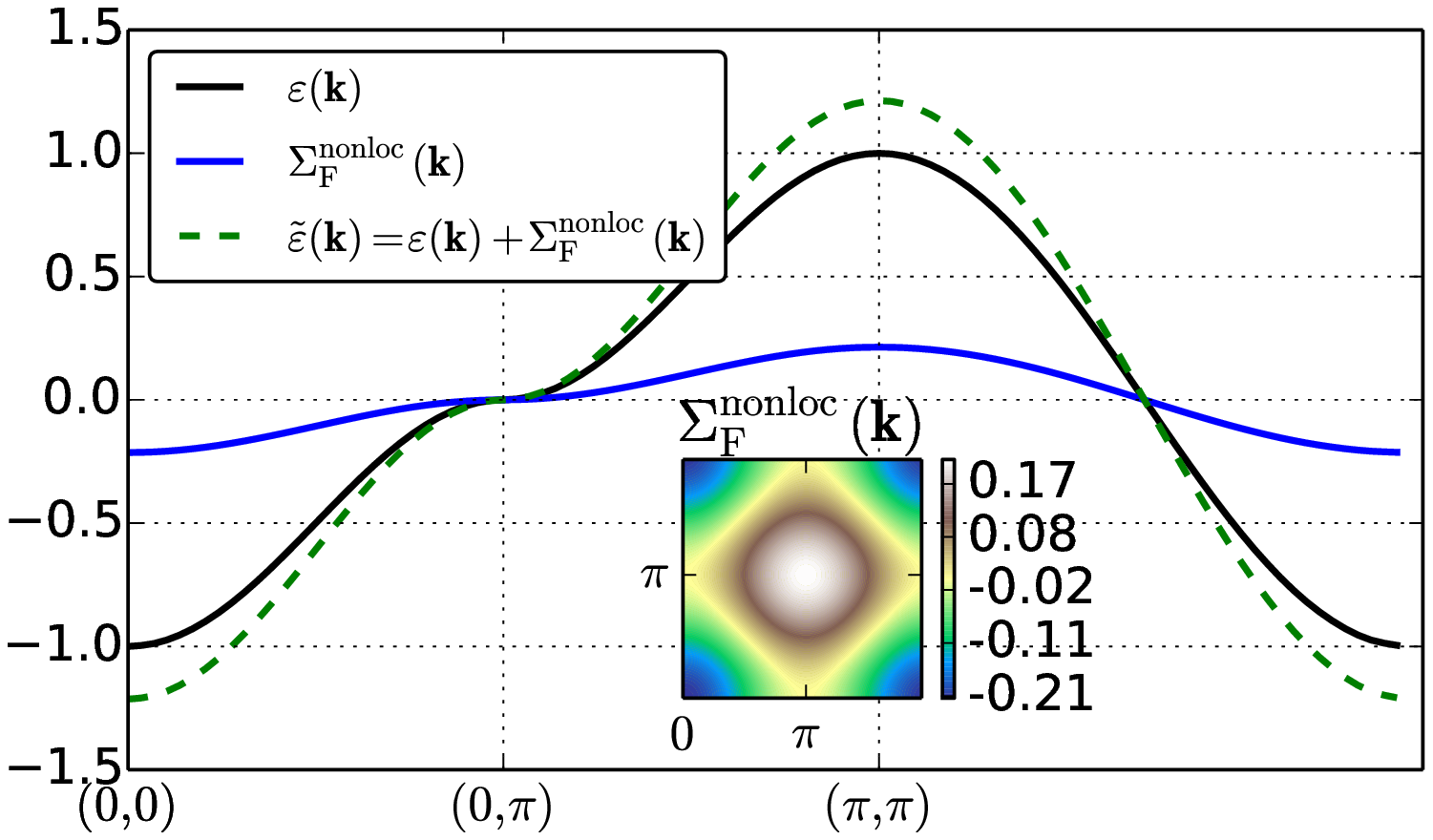}
\caption{(color online) Bare dispersion $\varepsilon(\vec{k})$ (solid black lines), nonlocal Fock self-energy $\Sigma_{\mathrm{F}}^{\mathrm{nonloc}}(\vec{k})$ (solid blue line) and renormalized dispersion $\tilde{\varepsilon}(\vec{k})$ (dashed green line) along a high-symmetry path in the Brillouin zone in the $GW$+EDMFT method. \emph{Inset}: color plot of $\Sigma_{\mathrm{F}}^{\mathrm{nonloc}}(\vec{k})$ in the first Brillouin zone. ($U=2.0$ and $V=0.4$, $\beta=100$, half-filling.) 
\label{fig:dispersions}}
\end{figure}

The nonlocal Fock term of Eq.~\eqref{eq:Sigma_expr}, which is real-valued and instantaneous, renormalizes the bandwidth. It can become quite large and momentum-dependent.
This is illustrated in Fig.~\ref{fig:dispersions} for the parameters $U=2$, $V=0.4$ (and $t'=0$, $V'=0$, which is assumed in the following if not explicitly stated otherwise).
The figure also indicates that for the case of nearest-neighbor hopping and interaction, the nonlocal Fock term can be exactly absorbed into the bare dispersion Eq.~\eqref{eq:epsk} by defining a $U$- and $V$-dependent hopping
\begin{equation}
\tilde{t}(t',U,V,V')=t+\delta t(t',U,V,V').\label{eq:renormalized_hopping}
\end{equation}
This can be understood by looking at the real-space representation of the Fock term, Eq.~\eqref{eq:Sigma_Gv},
\begin{align}
\Sigma_{\mathrm{F}}^{\mathrm{nonloc}}{}_{ij}
= & -G_{ij}(\zeroplus)v_{ij} + G_{ii}(\zeroplus)v_{ii}\delta_{ij}\nonumber \\
  = & -G_{\left<ij\right>}(\zeroplus)V\delta_{\left<ij\right>}\nonumber\\
    &-G_{\left<\left<ij\right>\right>}(\zeroplus)V'\delta_{\left<\left<ij\right>\right>},\label{eq:explicit_sigmann}
\end{align}
where the notation $\left<ij\right>$ [$\left<\left<ij\right>\right>$] denotes
a restriction to nearest-neighbor [next-nearest-neighbor] interactions.

Thus, for the case of Fig.~\ref{fig:dispersions}, where $V'=0$, the Fock term enters Eq.~\eqref{eq:G} as a renormalization of the nearest-neighbor hopping by
\begin{equation}
  \label{eq:deltat}
  \delta t(t',U,V,V')=-G_{\left<ij\right>}(\zeroplus)V .
\end{equation}
The Green's function term, which is closely related to the occupation number, has an implicit dependence on all parameters of the lattice problem.
In the presence of a next-nearest-neighbor interaction, $t'$ also gets renormalized according to
\begin{equation}
  \label{eq:deltatprime}
  \delta t'(t',U,V,V')=-G_{\left<\left<ij\right>\right>}(\zeroplus)V' .
\end{equation}
Note that such a term breaks the particle-hole symmetry of the lattice. In the particle-hole symmetric case with $t'=0$ and half filling, the next-nearest-neighbor occupation term vanishes: $G_{\left<\left<ij\right>\right>}(\zeroplus)=0$.

\section{Effective bandstructure}
\label{sec:theeffectivebandstructure}

\begin{figure}
  \includegraphics[width=1\columnwidth]{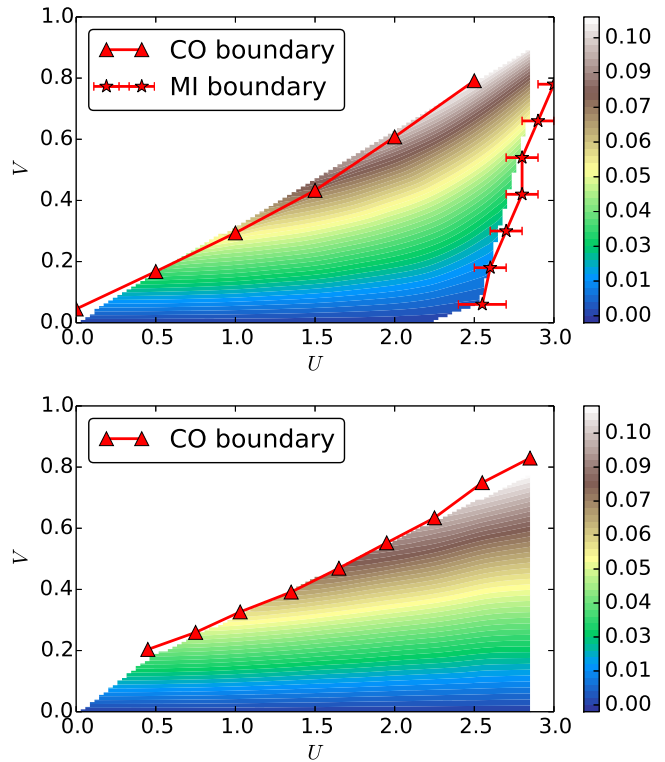}
  \caption{Hopping renormalization $\delta t$ according to equation~\eqref{eq:deltat} for half-filling (top panel) and $n=0.8$ (bottom panel).  The solid line with triangles indicates the phase transition to the charge-ordered phase and the line with stars the transition to the Mott insulating phase. 
    \label{fig:colorplot_fock_widening_half_filling}}
\end{figure}

In the simplest case of nearest-neighbor interaction $V$, the nearest-neighbor hopping renormalization $\delta t$ determines the band widening (see Eq.~\eqref{eq:deltat}). Since the half bandwidth is $D=4|t|$, the widening will be $\delta D=4\delta t$.
Figure~\ref{fig:colorplot_fock_widening_half_filling} illustrates this effect throughout the homogeneous part of the phase diagram for the particle-hole symmetric ($t'=0$) half filled case.
The most obvious feature is the increase with $V$, that is expected from Eq.~\eqref{eq:deltat} and the decrease close to the Mott-insulating phase.
$\delta t$ nonetheless remains significant even at very high values of $U$, a property that will be further investigated in Section~\ref{sec:insidethemottphase}.

Away from half filling, where the Mott-insulating phase does not exist, the corresponding suppression of $\delta t$ disappears, but otherwise the dependence on $V$ and $U$ is very similar to the half-filled case, see bottom panel of Fig.~\ref{fig:colorplot_fock_widening_half_filling}.

\begin{figure}
  \includegraphics[width=1\columnwidth]{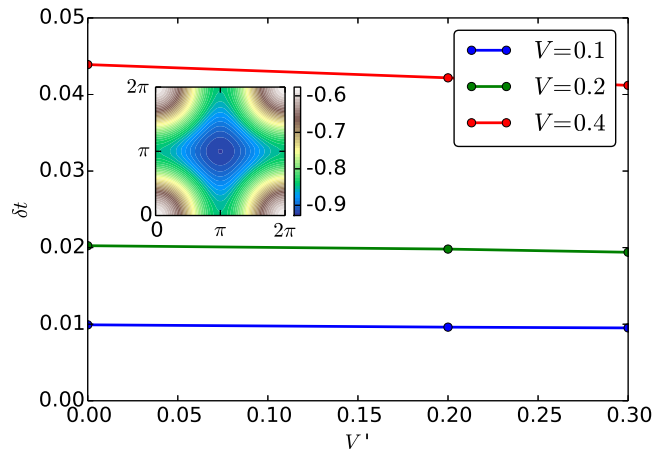}
  \caption{Dependence of the nearest-neighbor hopping parameter renormalization $\delta t$ on the next-nearest neighbor interaction $V'$ for fixed $U=2.0$, $t'=1/\sqrt{2}$ and $\left<n\right>=0.8$. \textit{Inset:} color plot of $\Sigma_{\mathrm{F}}(\vec{k})$ in a calculation with $32\times 32$ $\vec{k}$-points 
  for the $V=0.4$, $V'=0.2$ case.
}
  \label{fig:deltat_vs_Vp}
\end{figure}
To study a model with broken particle-hole symmetry, we introduce a nearest-neighbor hopping $t'=t/\sqrt{2}$ and fix the filling at $\left<n\right>=0.8$ as well as the nearest-neighbor interaction $V$. We then calculate $\delta t$ as a function of the next nearest nearest-neighbor interaction $V'$. As shown in Fig.~\ref{fig:deltat_vs_Vp}, 
the main effect on the hopping renormalization comes from the (essentially linear) dependence on $V$.
The qualitative effect of $V'$ is to slightly \emph{reduce} the renormalization.

In order to make the connection to realistic electronic structure
calculations, we note as a side remark that there
the situation is slightly more subtle. The band-widening effect is indeed relative to the reference point. Let us consider three reference Hamiltonians, 

(i) the Kohn-Sham Hamiltonian $H_\mathrm{KS}$
of Density Functional Theory (DFT),

(ii) the Hartree Hamiltonian $H_0 = H_\mathrm{KS} - V[v_\mathrm{xc}(\vec{r})]$, where $v_\mathrm{xc}(\vec{r})$ denotes the Kohn-Sham exchange-correlation potential, which is local in the electronic structure sense
(that is, ``local'' denotes a function depending only on one space variable $f(\vec{r})$, while ``nonlocal'' denotes a function depending on two variables $f(\vec{r},\vec{r'})$),

(iii) the nonlocal exchange Hamiltonian $H_\mathrm{xc}^\mathrm{F} = H_0 + V[v_\mathrm{xc}^\mathrm{F}(\vec{r},\vec{r'})]$ (where $v_\mathrm{xc}^\mathrm{F}(\vec{r},\vec{r'})$ denotes an exchange-correlation potential including ``nonlocal'' Fock exchange).

Then the hierarchy of the bandwidths in a metallic system is
\begin{equation}
  H_0 > H_\mathrm{xc}^\mathrm{F} > H_\mathrm{KS}.
\end{equation}

Thus, $H_\mathrm{xc}^\mathrm{F}$ indeed widens the band \emph{with respect to density functional theory calculations}. The question of the relative bandwidth changes
thus implies a question on the starting band structure.
We refer the interested reader to Ref.~\onlinecite{Hirayama2015} 
for a systematic construction of explicit low-energy many-body
Hamiltonians.

Here, we only comment on the specific point of the Hartree and Fock
terms, in order to put our work on the extended Hubbard Hamiltonian
into perspective with respect to realistic electronic structure calculations.
Indeed, as argued in Ref.~\onlinecite{Hirayama2015}, in realistic 
electronic structure calculations one needs to avoid double counting 
of interactions at the one- and two-particle level. 
Let us consider first the case of the Hartree terms: standard electronic
structure techniques (e.g. a DFT calculation) produce a band structure
including the Hartree contribution. This one-body potential contribution
is then already part of the effective hopping parameter determined
from this band structure.
Ref.~\onlinecite{Hirayama2015} explains how to avoid double counting 
by including -- at the level of the many-body calculation -- only terms 
beyond Hartree. Here, we do not need to address this point in detail, since
a Hartree term included in the model calculation would cancel out
with the corresponding shift of the chemical potential, since the
particle number is eventually determining the energetic level of the
single-orbital included in the present model.

Let us now move to the analogous question for the Fock term: one may examine the relevance of excluding it at the level of the many-body calculation, and keeping it at the level of the electronic structure calculation instead.
The answer is based on several elements: The first point to note
is that standard DFT calculations do treat exchange in a local
approximation (where ``local'' here means again ``local in the electronic 
structure sense'', see above), which
relies on an error cancellation effect with part of the correlation
contribution (see e.g. Ref.~\onlinecite{VanRoekeghem2014a}) and is not relevant here. 
The next question is therefore: why not start from a
Hartree-Fock calculation in the continuum in the full energy range
of the Coulomb Hamiltonian? Such a treatment would neglect the 
crucial screening of the bare interaction by high-energy degrees of
freedom (typically, matrix elements of the bare Coulomb interaction
in the relevant Wannier functions are of the order of several tens
of electron volts, while the effective Hubbard interactions are usually a few electron volts).
Therefore, what is relevant here is indeed the exchange term calculated
using the effective bare interaction of the low-energy Hamiltonian.
For realistic electronic structure calculations, this interaction
should correspond to a partially screened interaction, where screening
by high-energy degrees of freedom is taken into account (as done e.g.
in the screened exchange + DMFT scheme 
\cite{VanRoekeghem2014, VanRoekeghem2016b}).
We refer the interested reader to Refs.~\onlinecite{Hirayama2015, VanRoekeghem2014, VanRoekeghem2014a} for details.

\section{Simplified variants of $GW$+EDMFT}
\label{sec:withandwithoutfock}
\begin{figure}
\includegraphics[width=1.0\columnwidth]{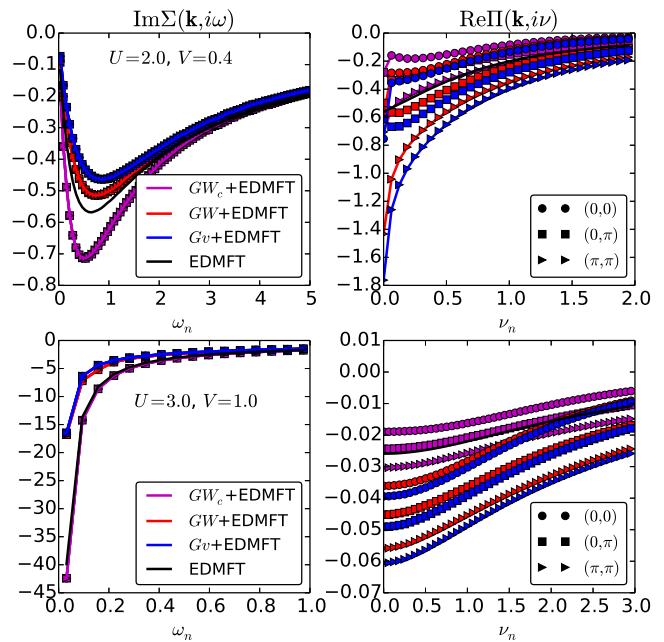}
\caption{(color online) $\mathrm{Im}\Sigma(\vec{k},i\omega)$ (left) and $\mathrm{Re}\Pi(\vec{q},i\nu)$ (right) (black lines for EDMFT, magenta lines for $GWc$+EDMFT, red lines for $GW$+EDMFT, blue lines for $Gv$+EDMFT). Circles, squares and triangles show results for the $(0,0)$, $(0,\pi)$ and $(\pi,\pi)$ points, respectively. (Top: $U=2.0$ and $V=0.4$; bottom: $U=3.0$ and $V=1.0$;$\beta=100$; $n=1$.)\label{fig:nonlocal}}
\end{figure}
In the following, we study the effect of this $V$-dependent bandwidth renormalization on local observables as well as the critical value of the nearest-neighbor repulsion for the transition into the charge-ordered phase.
We will call ``$GW_{c}$+EDMFT'' the formula implemented in Ref.~\onlinecite{Ayral2013} (which contains only the $GW_c$
term, see Ref.~\onlinecite{Ayral2016a}), and ``$GW$+EDMFT'' the formula with the self-energy expression
(\ref{eq:Sigma_expr}). For comparison, we also show results for ``$Gv$+EDMFT'',
 a scheme where $\Sigma$ is the sum of the impurity self-energy
and of the nonlocal Fock term only (the first and third terms of Eq.~(\ref{eq:Sigma_expr})). In all three schemes, the polarization is
the sum of the impurity polarization with the nonlocal part of the
interacting bubble, as described in Ref.~\onlinecite{Ayral2013},
steps (5) (a) and (5) (b) of section V.

\begin{figure}
\begin{centering}
\includegraphics[width=1\columnwidth]{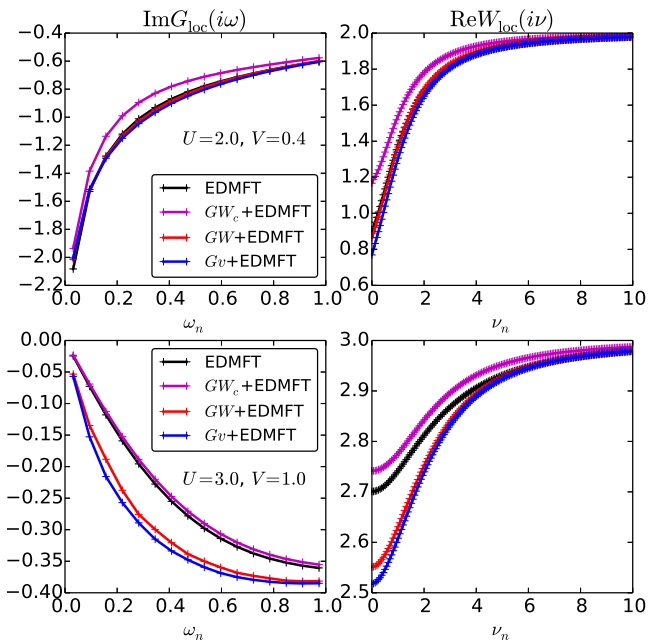}
\par\end{centering}
\caption{(color online) $\mathrm{Im}G_{\mathrm{loc}}(i\omega)$ (left) and
$\mathrm{Re}W_{\mathrm{loc}}(i\nu)$ (right) (black lines for EDMFT,
magenta lines for $GW_{c}$+EDMFT, red lines for $GW$+EDMFT, blue lines
for $Gv$+EDMFT). 
The top panels are for $U=2.0$ and $V=0.4$, and the bottom panels for $U=3.0$ and $V=1.0$ ($\beta=100$, $n=1$).
\label{fig:prb}}
\end{figure}

\begin{figure}
\begin{centering}
\includegraphics[width=1\columnwidth]{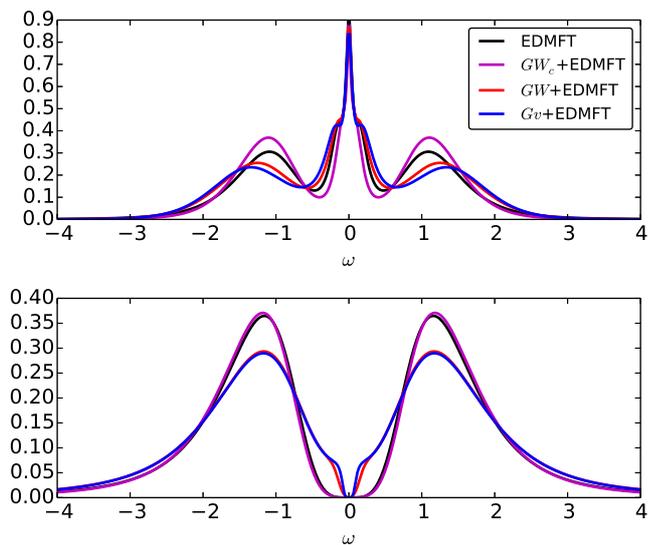}
\par\end{centering}
\caption{(color online) Local spectral functions obtained by MaxEnt\cite{Bryan1990,Jarrell1996}
analytical continuation (black for EDMFT, red for $GW_{c}$+EDMFT,
magenta for $GW$+EDMFT, blue lines for $Gv$+EDMFT). The top panels are for $U=2.0$ and $V=0.4$, and the bottom panels for $U=3.0$ and $V=1.0$ ($\beta=100$, $n=1$).\label{fig:spectral}}
\end{figure}

In Fig.~\ref{fig:nonlocal}, we plot the self-energy and polarization
obtained from the three schemes at different momenta, and we compare
the results to the local EDMFT self-energy and
polarization. All these results are for half-filling. One can observe the following
trends:

(i) While the imaginary part of the self-energy in $GW_c$+EDMFT 
is larger than in EDMFT, the opposite is true for $GW$+EDMFT, i.e.,
the $GW$+EDMFT self-energy is less correlated than the EDMFT
self-energy. 

(ii) The $GW$+EDMFT result is more strongly correlated than $Gv$+EDMFT. 

(iii) At small Matsubara frequencies, the polarization is overall larger
in the $GW$/$Gv$+EDMFT method than in the $GW_c$+EDMFT. 

The trend in the self-energy (i.e. less correlated in $GW$+EDMFT than
EDMFT) can be understood easily from the broadening effect of the
nonlocal Fock term on the band: when the (effective) bandwidth gets
larger, so does the polarization $P$, and hence screening effects
are more important, interactions are more screened and as a result,
the imaginary part of the Matsubara self-energy is smaller in absolute
magnitude.

\begin{figure}[ht]
\begin{centering}
\includegraphics[width=1\columnwidth]{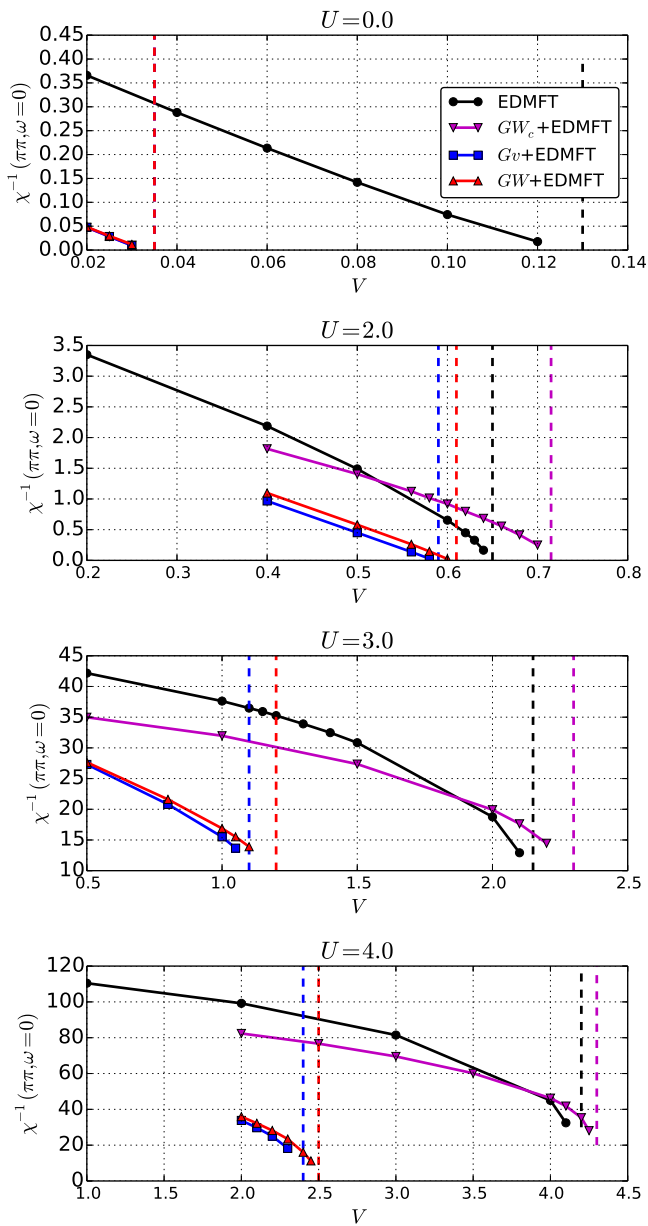}
\par\end{centering}
\caption{(color online) $\chi^{-1}(\pi\pi,\omega=0)$ as a function of $V$
for (from top to bottom) $U=0$, $U=2$ , $U=3$ and $U=4$ (black lines
for EDMFT, magenta lines for $GW_c$+EDMFT, red lines for $GW$+EDMFT, blue
lines for $Gv$+EDMFT). The dashed lines are an estimate of the critical
$V_{c}$ ($\beta=100$). \label{fig:phase_diag_construction}
}
\end{figure}

Less trivial is the comparison between $GW$+EDMFT and $Gv$+EDMFT. Here, the band-widening effect
is included in both calculations, and it turns out that the additional nonlocal $GW$ contributions to the self-energy lead to stronger correlations.
This is consistent with the conclusions of  Ref.~\onlinecite{Ayral2013},
which compared $GW_c$+EDMFT to EDMFT. 

In the top panels of Fig.~\ref{fig:prb}, we replot panels (a) and (b) of Fig. 15 of Ref.~\onlinecite{Ayral2013}, and we show, in the bottom panels, the same observables for $U=3$, $V=1$. 
One sees that the deviation of the $GW$+EDMFT and $Gv$+EDMFT results from 
the EDMFT result is very small for $U=2$, $V=0.4$, but sizable for larger interaction values
($U=3$, $V=1$).

\begin{figure}
\begin{centering}
\includegraphics[width=1\columnwidth]{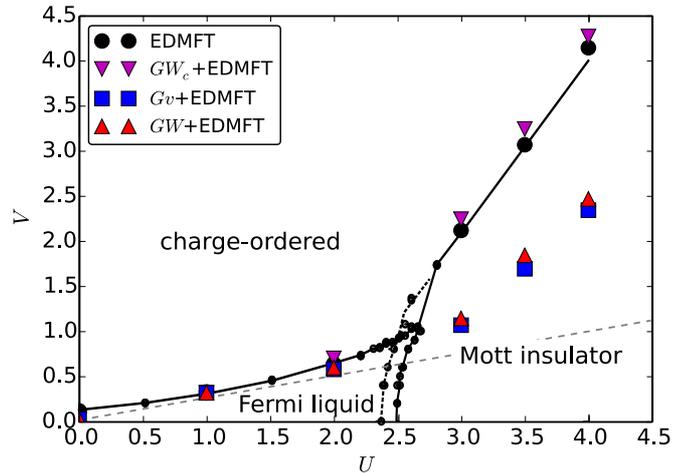} 
\par\end{centering}
\caption{(color online) EDMFT, $GW$+EDMFT and $Gv$+EDMFT phase diagrams. The black
lines are EDMFT results reproduced from Fig.~5 of Ref.~\onlinecite{Ayral2013}.
The $GW_{c}$+EDMFT phase boundaries, magenta triangles, are very close to the EDMFT
phase boundaries. Red triangles show the $GW$+EDMFT result, blue squares to the $Gv$+EDMFT result. The dashed grey line corresponds
to the interaction energy estimate of $V_{c}$, $V_{c}^{\mathrm{int}}=U/4$ (see text).\label{fig:EDMFT-and-GW+EDMFT}
}
\end{figure}

In Fig.~\ref{fig:spectral}, we 
plot the corresponding spectral functions (the EDMFT and $GW_c$+EDMFT results are identical to Fig.~2(b) of Ref.~\onlinecite{Ayral2012}).
As a logical consequence of the above observations, the $GW$+EDMFT and
$Gv$+EDMFT spectra are close to each other and slightly less correlated
than the EDMFT spectrum, in the sense that the integrated weight of
the quasiparticle peak is larger in those methods.

We next consider the phase diagram in the $U$-$V$ plane.
In Fig.~\ref{fig:phase_diag_construction}, we plot the dependence
of the inverse charge susceptibility $\chi^{-1}_{\vec{q}=\pi,\pi}(i\nu_m=0)$ on the nearest-neighbor repulsion $V$, with $\chi$ defined by
\begin{equation}
  \label{eq:chicq}
  \chi_\vec{q}(i\nu_m)=\frac{-\Pi_\vec{q}(i\nu_m)}{1-\Pi_\vec{q}(i\nu_m)v_\vec{q}}.
\end{equation}
When the inverse susceptibility vanishes, the
charge susceptibility diverges, signaling a transition to a charge-ordered
phase with a checkerboard pattern. The corresponding phase diagram is shown in Fig.~\ref{fig:EDMFT-and-GW+EDMFT}, where we plot the results from Fig.~5 of Ref.~\onlinecite{Ayral2013} together with the phase boundaries for the $GW$+EDMFT and $Gv$+EDMFT methods.

At low and intermediate $U$, $GW$+EDMFT and $GW_c$+EDMFT yield quantitatively similar critical nonlocal interactions $V_{c}$ for the transition to the charge-ordered phase over a wide range of the local interactions. More importantly, they capture the expected $GW$ behavior at low $U$ that EDMFT misses due to its local self-energy.
In the strong-coupling limit, the value of $V_c$ is substantially reduced (middle and bottom panels) when going from EDMFT to $GW$+EDMFT or even only
$Gv$+EDMFT ($GW_c$+EDMFT is very close to EDMFT).

\section{Inside the Mott phase}
\label{sec:insidethemottphase}
In the large-interaction regime of the phase diagram (Fig.~\ref{fig:EDMFT-and-GW+EDMFT}), the nonlocal Fock term has a significant effect. The schemes which lack this instantaneous contribution, EDMFT and $GW_c$+EDMFT, yield a larger and steeper phase boundary than the schemes that take the Fock term into account ($Gv$+EDMFT and $GW$+EDMFT).

The exact phase boundary in the Mott phase is difficult to predict \emph{a priori}. 
It can be computed in the classical ($t_{ij}\rightarrow 0$) and zero-temperature limit of the extended Hubbard model by exact Monte-Carlo simulations, as e.g. in Ref.~\onlinecite{Pawlowski2006}, and is given by the analytical expression 
\begin{equation}
V_c^\mathrm{int}=U/4,
\label{eq:classical_V}
\end{equation} 
where 4 corresponds to the number of nearest neighbors.
This line is plotted as a dashed grey line in Fig.~\ref{fig:EDMFT-and-GW+EDMFT}. This result can be obtained by a simple comparison between the interaction energies of the Mott-insulating phase and of the checkerboard phase. In the full-fledged model, finite temperature and quantum tunneling have to be taken into account. In the low-temperature regime ($T=0.01$) of Fig.~\ref{fig:EDMFT-and-GW+EDMFT}, the deviation between the classical solution and the solution to the full quantum problem comes mostly from the quantum tunneling kinetic term. 

To guess the influence of the quantum tunneling term, one may observe that the effect of temperature in the classical problem is to enhance the value of $V_c$, i.e to disfavor the charge-ordered phase over the Mott phase.\cite{Pawlowski2006} Since the quantum tunneling (hopping) has a physical effect similar to temperature in classical systems,\cite{Biermann1998}  
namely to delocalize the particles 
one may speculate that it will also lead to a higher $V_c$ in the quantum case. 

In fact, this feature is present by construction in the EDMFT and $GW$+EDMFT schemes. The denominator in the susceptibility (Eq.~\eqref{eq:chicq}) imposes that the charge-ordering transition should occur for negative values of $v_\vec{q}$, since the polarization $\Pi_\vec{q}$ is always negative (for the parameters studied here). For the square lattice, this implies $V_c>\frac{U}{4}$, which is the classical energy estimate (Eq.~\eqref{eq:classical_V}). Therefore, by construction, in $GW$+EDMFT schemes, introducing hopping on the lattice will always favor the disordered phase. 
This is indeed what is seen in all variants. We also observe that the method including most diagrams, $GW$+EDMFT, has a phase boundary which is much closer to the classical limit than the comparatively cruder EDMFT approximation.

In order to gain a better qualitative understanding of the large-$U$ behavior, we have performed an analytical self-consistent estimation of the value of the band-widening effect $\delta t$ (defined in Eq.~\eqref{eq:deltat}) coming from the Fock term. Approximating the self-energy as the sum of the atomic limit (in the spirit of the Hubbard-I approximation\cite{Hubbard1963}) and of the Fock self-energy, as described in more detail in Appendix~\ref{sec:Hubbard-I}, we obtain
\begin{equation}
  \delta t=\frac{tV}{2U-V}.
\end{equation}
Thus $\delta t$ may become arbitrarily
large if the nonlocal interaction coefficient exceeds twice the value of the local
one. However, even disregarding the fact that the generic case is
certainly the opposite one (local interactions in general exceed nonlocal
ones), one should be aware of the fact that in that case the Hubbard
I approximation, which is justified in the strong coupling limit,
would no longer be appropriate.

By inspecting the phase diagram in Fig.~\ref{fig:EDMFT-and-GW+EDMFT}, we can parametrize the phase boundary in the large-$U$ limit as a constant slope, i.e   $V_c=\alpha U+\beta$. [Within the $U$-range that we can simulate (we performed measurements up to $U=8.0$), we can estimate $\alpha=1.25$ and $\beta=-2.5$.] With this parametrization,
we obtain 
\begin{equation}
  \delta t_c=\frac{t(\alpha U+\beta)}{2U-\alpha U-\beta}.
\end{equation}
Hence, the bandwidth 
renormalization (proportional to $\delta t$) 
stays relevant in the vicinity of the charge-ordering transition 
even at large $U$.

\section{Beyond $GW$+EDMFT: Comparison to dual bosons and TRILEX}
\label{sec:DB}

As mentioned in the introduction, the EDMFT formalism suffers from certain conceptual problems, such as the lack of thermodynamic consistency and an unreliable description of collective modes. These shortcomings are alleviated in the recently developed dual boson (DB) method,\cite{Rubtsov2011} which, in its full-fledged implementation,\cite{VanLoon2014, Stepanov2015} computes the susceptibility\cite{Hafermann2014,vanLoon2015} after resumming an infinite number of ladder diagrams built from local impurity four-leg vertices. 

These four-leg vertices, which are also central to the Dynamical Vertex Approximation\cite{Toschi2007, Katanin2009, Schafer2013, Valli2014, Schafer2015, Schafer2016} (which was recently shown to be a simplified version of QUADRILEX, a method consisting in an atomic approximation of the four-particle irreducible functional\cite{Ayral2016}), 
can nonetheless only be obtained at a considerable computational expense and require a proper parametrization and treatment of their asymptotic behavior.\cite{Kunes2011, Boehnke2011, Li2015a, Wentzell2016} Consequently, it is not
possible to use them routinely in
multi-orbital calculations (see however \cite{Boehnke2014, Galler2016}) 
and lightweight improvements on EDMFT, especially with realistic applications in mind, are desirable. Recent attempts to forgo the computation of four-leg vertices include the TRILEX method\cite{Ayral2015, Ayral2015c} and simplified dual boson schemes such as DB+$GW$ or DB+$GW\gamma$.\cite{Stepanov2016} Whether they retain the abovementioned conserving properties, however, is yet unclear.

In fact, the results obtained in the simplified ``dual'' approaches that include at least the electron-boson vertex $\gamma$ are similar to those obtained by the $GW$+EDMFT method, which is conceptually and practically simpler than dual methods, and has hence already been applied to realistic materials in a number of works.\cite{Tomczak2012, Hansmann2013, Hansmann2016, Boehnke2016} 

\begin{figure}
\includegraphics[width=1.0\columnwidth]{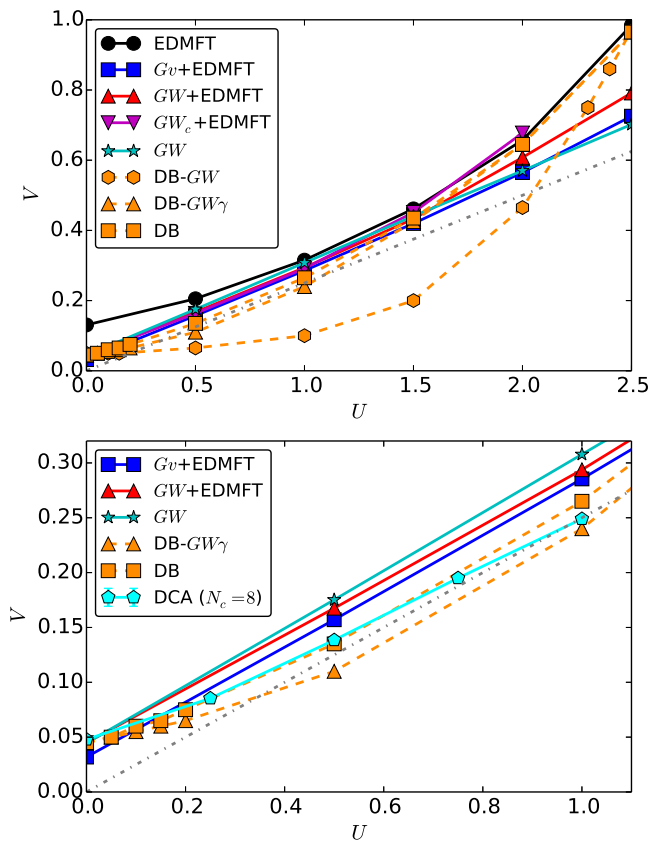}
\caption{(color online) Comparison of the $U$-$V$ phase diagrams with results from Ref.~\onlinecite{Stepanov2016} 
using different variants of the dual boson scheme
(orange lines). The dashed grey line is the 
mean field estimate $V=U/4$. 
Lower panel: zoom on the low-$U$ region, and additional comparison to the DCA results of Ref.~\onlinecite{Terletska2016} (cyan pentagons; $N_c=8$, obtained via a linear extrapolation\cite{Terletska2016}). For all results, $\beta = 50$. \label{fig:stepanov_comparison}}
\end{figure}
In Fig.~\ref{fig:stepanov_comparison}, we compare the phase diagram for model (\ref{eq:eHubbard}) obtained from various simplified variants\cite{Stepanov2016} of the dual boson scheme, and compare it to $GW_c$+EDMFT, $GW$+EDMFT, and the $GW$ approximation. We restrict this comparison to values of $U$ below the Mott transition, for lack of available dual-boson results in the Mott-insulating phase.

Let us start with the small-$U$ limit.
For $U<1.5$, the phase boundaries obtained in all the $GW$+EDMFT as well as $GW$ alone are almost indistinguishable.
The $GW$ transition is a straight line for all shown values, the variants with an impurity polarization have varying degrees of up-curvature, with the $GW$+EDMFT line in between the $GW_c$+EDMFT and the $Gv$+EDMFT line.
For $Gv$+EDMFT it is notable that the $U\rightarrow 0$ limit does not reproduce the $GW$ value.  

The dual boson lines start with a similar upwards trend, with the exception of the DB-$GW$ line, which follows essentially the HS-$V$ variant of 
$GW$+EDMFT 
(more properly denoted as $GD$~+ second order perturbation theory (SOPT)~+ EDMFT)
as shown in Ref.~\onlinecite{Stepanov2016}, 
and discussed in detail at the end of this section. 
Yet, the dual-boson variants start out with a lower slope, indicating stronger ordering tendencies already for the lowest values of $U$, while all $GW$+EDMFT variants follow the slope of the `weak-coupling' $GW$ boundary in the vicinity of $U=0$.\footnote{Note that   the dual-boson variants and HS-$V$ calculations of Ref.~\onlinecite{Stepanov2016} have been executed as a single lattice self-consistency iteration on top of the converged EDMFT solution.} 
Interestingly, a very recent cluster-EDMFT study \cite{Terletska2016} 
 reports a similarly reduced slope in the weak-$U$ regime. 
 We also note that only the full DB critical line is above the $U/4$ line (dashed grey line, discussed in section~\ref{sec:insidethemottphase}), while the (non-self-consistent) DB-$GW$ and DB-$GW\gamma$ results are not (the latter only slightly so). Further comparisons of DB with $GW$+EDMFT (in the HS-$V$ decoupling) can be found in Ref.~\onlinecite{Stepanov2016}, see Fig. 8.

For $U>2$, the phase boundary for $GW$+EDMFT is below the dual-boson phase boundary, while $GW$ is even a bit lower. 
For $GW_c$+EDMFT, the $U=2.5$ point already falls in the Mott-insulating phase and is not shown in the comparison. The too low Mott-transition line for $GW_c$+EDMFT is not surprising, since it lacks the band-widening of the nonlocal Fock term. 

Two possible decouplings were previously discussed in the literature, ``HS-$UV$'' 
(giving rise to $GW$+EDMFT) and ``HS-$V$'' (
resulting in a combined ``$GD$+SOPT+EDMFT'' scheme, 
where $D$ is the screened non-local interaction, 
see Ref.~\onlinecite{Ayral2013}).
We emphasize that, contrary to the HS-$UV$ variant and e.g. the random-phase approximation (RPA), the HS-$V$ variants do not resum, in the nonlocal part of the self-energy, the local ($U$) and nonlocal ($V$) parts of the interaction to the same order. (They are resummed, respectively, to second and infinite order.) This arbitrary inconsistency raises questions concerning the soundness of the HS-$V$ scheme, as already pointed out in Ref.~\onlinecite{Ayral2013}.

Recent works have indeed confirmed the deficiency of ``HS-$V$''.
For instance, all the 
``EDMFT+$GW$''-related variants shown in Fig.~7 of Ref.~\onlinecite{Stepanov2016}, which were obtained in a HS-$V$ flavor, yield phase boundaries which are much lower than either DB or the $GW$+EDMFT results in Fig.~\ref{fig:stepanov_comparison} (which correspond to the HS-$UV$ variant of the decoupling of the interaction). The only additional outlier is the simplest DB type of approximation, the DB-$GW$ variant, for which Ref.~\onlinecite{Stepanov2016} showed that it is formally similar to a HS-$V$ calculation.  
More recently, Ref.~\onlinecite{Terletska2016} used cluster dynamical mean field theory to study the extended Hubbard model, which allows a \emph{control} on errors by increasing the size of the cluster (but neglecting inter-cluster interactions, which in the case of EDMFT and $GW$+EDMFT are treated via the retarded impurity interactions). These cluster results were shown to be in poor agreement with ``HS-$V$'', but very close to the full-fledged DB method. In the lower panel of Fig.~\ref{fig:stepanov_comparison}, we show that the $GW$+EDMFT (HS-$UV$) method yields a critical $V_c$ in agreement (with a 20\% accuracy or better) with the cluster results, a remarkable result in view of the reduced numerical cost of this method compared to cluster DMFT. Comparisons for the larger $U$ values in Fig.~\ref{fig:stepanov_comparison} would be of great interest. 

\begin{figure}
\includegraphics[width=1.0\columnwidth]{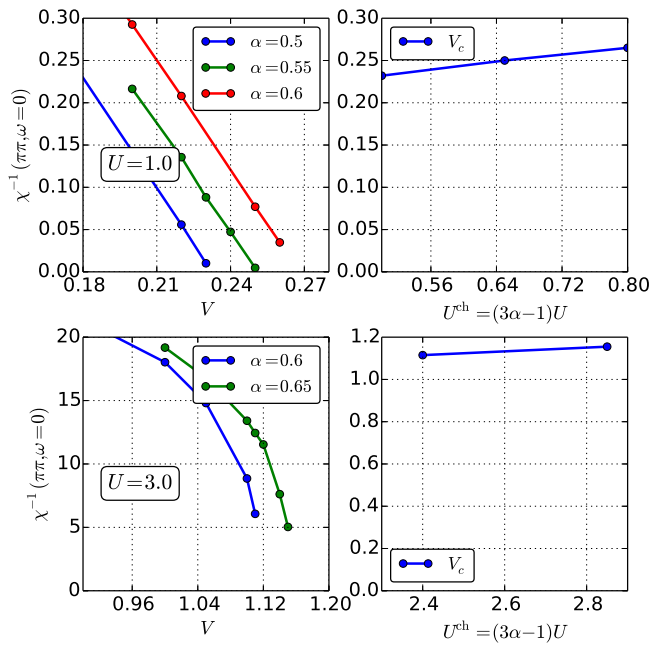}
\caption{(color online) Inverse static charge susceptibility at $Q=(\pi,\pi)$ (left column) and dependence of the critical $V$ on the decoupling (right column) for $U=1$ (top row) and $U=3$ (bottom row) within single-site TRILEX.\label{fig:trilex}}
\end{figure}
We end this section by examining two further questions, namely the influence of spin fluctuations and the impact of local vertex corrections. One can expect that neither are important for the charge-ordering instability under study, since (i) this is an instability in the charge channel, not the spin channel, and (ii) as $V$ increases towards charge ordering, the effective static interaction $\mathcal{U}(\omega=0)$ decreases to zero,\cite{Ayral2013} making the system behave more and more like a weakly-correlated metal, where vertex corrections are expected to be small.

In all previous implementations of the $GW$+EDMFT method, the interaction was formally decoupled in the charge-channel only, neglecting the possible influence of spin fluctuations. Furthermore, in $GW$+EDMFT, the influence of the local vertex on the nonlocal self-energy is included only through the nonlocal Green's function. 
In the TRILEX approximation, both charge and spin fluctuations are taken into account, as well as local vertex corrections to the nonlocal self-energy.

We can thus answer both questions of interest by implementing the TRILEX method for the extended Hubbard model. We refer the reader to Refs.~\onlinecite{Ayral2015,Ayral2015c} for implementation details. The only difference with respect to the application to the Hubbard model is that Eq.~(41) of Ref.~\onlinecite{Ayral2015c} must be modified to also describe nonlocal interactions, which means that Eq.~(61b) of that publication becomes
\begin{equation}
  W^\eta(\mathbf{q},i\Omega) = \frac{v^\eta(\mathbf{q})}{1 - v^\eta(\mathbf{q})P^\eta(\mathbf{q}, i\Omega) },
\end{equation}
 where $\eta$ denotes the charge ($\mathrm{ch}$) or spin ($\mathrm{sp}$) channel and
\begin{eqnarray}
  v^\mathrm{ch}(\mathbf{q})&=&U^\mathrm{ch} + 2 V(\cos(q_x)+\cos(q_y)),\\
  v^\mathrm{sp}(\mathbf{q})&=&U^\mathrm{sp},
\end{eqnarray}
and the bare on-site interactions in the charge and spin channels are parametrized, in the so-called Heisenberg decoupling,\cite{Ayral2015c} by a parameter $\alpha$:

\begin{equation}
  U^{\mathrm{ch}}=(3\alpha-1)U,\;\;U^{\mathrm{sp}}=(\alpha-2/3)U.
\end{equation}

In Fig.~\ref{fig:trilex}, we show TRILEX results for two characteristic points of the phase diagram, namely $U=1$ (characteristic of the metallic phase), and $U=3$ (characteristic of the Mott phase). 
First, we observe that the critical $V_c$ (computed by looking for a vanishing inverse static susceptibility, shown in the left panels) is quite close to that of $GW$+EDMFT, justifying our a priori intuition. This agreement is quite remarkable, since $GW$+EDMFT has only charge fluctuations, while TRILEX has both charge and spin fluctuations.
Second, $V_c$ only mildly depends on the ratio of the charge to spin fluctuations, as can be seen in the right panels, where quite large variations of $U^\mathrm{ch}$ (and correspondingly $U^\mathrm{sp}$) lead to comparatively small variations in $V_c$. 

Taking inspiration from the comparison of the cluster extension of TRILEX with exact benchmark results for the two-dimensional Hubbard model\cite{Ayral2017c} (there, one observes that whenever the TRILEX solution is close to the exact solution, the dependence on the decoupling is weak), this stability (compared to charge-only $GW$+EDMFT, and with respect to $\alpha$) can be used as a proxy for the quantitative robustness of the present $GW$+EDMFT results.

\section{Conclusion}
In conclusion, we have shown that the nonlocal Fock term
has a significant influence on the description of the charge fluctuations
in the $GW$+EDMFT method, especially in the strong-coupling limit. By effectively enhancing the bandwidth, it
lowers the critical value of the nonlocal interaction for the charge-ordering transition.
We have also shown that the differences between the EDMFT and $GW$+EDMFT phase
diagrams are to a large extent a consequence of the nonlocal Fock
term, which is not included in EDMFT. 

Another interesting result is that the simple extension from EDMFT to a $Gv$+EDMFT
formalism yields results similar to the full-fledged $GW$+EDMFT method. This suggests the possibility of studying complex multiband materials, where a full $GW$+EDMFT computation would be too costly, using techniques in the spirit of the recent Screened exchange + dynamical DMFT (SEx+DMFT) 
method.\cite{VanRoekeghem2014, VanRoekeghem2014a, vanRoekeghem2016} 
In realistic materials, the simple single-band description is not
sufficient, and substantial screening effects resulting from the presence
of higher energy degrees of freedom must be taken into 
account.\cite{Aryasetiawan2004, Biermann2014, Werner2016} 

Performing a self-consistent calculation of the screening by these higher-energy states is however computationally expensive, even within a multi-tier approach,\cite{Boehnke2016}  where the updates are restricted to an intermediate energy window. A scheme which combines a properly renormalized bandstructure with a self-consistent treatment of screening effects within the low-energy subspace may provide a good basis for 
tractable, but still accurate {\it first principles}
electronic structure methods for correlated electron materials.

\acknowledgments

We acknowledge useful discussions with Y. Nomura, A. I. Lichtenstein, E. Stepanov
and E. van Loon. We thank A. Huber and A. I. Lichtenstein for providing us the dual-boson data for Fig.~\ref{fig:stepanov_comparison}, as well as H. Terletska for providing us the DCA data for Fig.~\ref{fig:stepanov_comparison}. 
This research was supported by SNSF through NCCR Marvel, 
IDRIS/GENCI Orsay (project number t2016091393),
ECOS-Sud MinCYT under project number A13E04, and the 
European Research Council (project number 617196). Part of the implementation is based on the TRIQS toolbox\cite{Parcollet2014} and on 
the ALPS libraries.\cite{Bauer2011a} 

\appendix
\section{Estimation of the bandwidth widening with combined Hubbard-I and Fock}
\label{sec:Hubbard-I}
In this appendix we discuss a simple Hubbard I (plus Fock) type treatment of the $U$-$V$ model.
These arguments are not meant to be exact or comprehensive, most notably we ignore the effect of the nearest-neighbor interaction on the local self-energy and any nonlocal screening, but they provide useful insights into the nontrivial nature of the large-$U$ and large-$V$ limit. 

We start with an approximation to the self-energy which follows the
spirit of the Hubbard-I approximation by taking the atomic $\frac{U^{2}}{4z}$ self-energy
locally, but goes beyond it by taking also the instantaneous nonlocal Fock contribution into account:
\[
\Sigma(\boldsymbol{k},z)=\frac{U^{2}}{4z}+2\delta t(\cos k_{x}+\cos k_{y})
\]

The corresponding Green's function reads:
\[
G(\boldsymbol{k},z)=\frac{1}{z-\tilde{\varepsilon}_{\boldsymbol{k}}-\frac{U^{2}}{4z}}
\]

with $\tilde{\varepsilon}_{\boldsymbol{k}}$ denoting the effective
dispersion including the Fock term:
\begin{equation}
\tilde{\varepsilon}_{\boldsymbol{k}}\equiv2(t+\delta t)(\cos k_{x}+\cos k_{y})\label{eq:vareps_def}
\end{equation}

Thus, we can write:

\begin{equation}
G(\boldsymbol{k},z)=\frac{z}{(z-z_{+}(\boldsymbol{k}))(z-z_{-}(\boldsymbol{k}))}\label{eq:G_interm}
\end{equation}

with
\begin{equation}
z_{\pm}(\boldsymbol{k})=\frac{\tilde{\varepsilon}_{\boldsymbol{k}}\pm\sqrt{\tilde{\varepsilon}_{\boldsymbol{k}}^{2}+U^{2}}}{2}.
\end{equation}
As expected, in the atomic limit ($\tilde{\varepsilon}_{\boldsymbol{k}}\rightarrow0$),
the function has two peaks at $\pm U/2$, corresponding to the two
Hubbard bands.

We can decompose the expression of Eq. (\ref{eq:G_interm}) as:

\begin{equation}
G(\boldsymbol{k},z)=\frac{A_{+}(\boldsymbol{k})}{z-z_{+}(\boldsymbol{k})}+\frac{A_{-}(\boldsymbol{k})}{z-z_{-}(\boldsymbol{k})}\label{eq:G_final}
\end{equation}

with
\begin{equation}
A_{\pm}(\boldsymbol{k})\equiv\frac{1}{2}\left(1\pm\frac{\tilde{\varepsilon}_{\boldsymbol{k}}}{\sqrt{\tilde{\varepsilon}_{\boldsymbol{k}}^{2}+U^{2}}}\right).
\end{equation}
$A_{+}$ and $A_{-}$ are the weights of the upper and lower Hubbard
bands, respectively. Using Eq. (\ref{eq:G_final}), one writes the
spectral function as:
\[
A(\boldsymbol{k},\omega)=A_{+}(\boldsymbol{k})\pi\delta(\omega-z_{+}(\boldsymbol{k}))+A_{-}(\boldsymbol{k})\pi\delta(\omega-z_{-}(\boldsymbol{k}))
\]

Under the assumption that the Hubbard bands are well separated ($U$
large enough), only the lower Hubbard band contributes to the occupancy
(at $T=0$ for simplicity):
\begin{equation}
n_{\boldsymbol{k}}=\int_{-\infty}^{0}\frac{d\omega}{\pi}A(\boldsymbol{k},\omega)\approx A_{-}(\boldsymbol{k})\approx\frac{1}{2}\left(1-\frac{\tilde{\varepsilon}_{\boldsymbol{k}}}{U}\right)\label{eq:occupancy_final}
\end{equation}

In the second equality, we have again used the fact that $U$ is large
enough (to neglect $\tilde{\varepsilon}_{\boldsymbol{k}}^{2}$ in
the square root).

On the other hand, the occupancy is related to $G(\boldsymbol{k},\tau=0^{+})$
in the following way:
\begin{equation}
n_{\boldsymbol{k}}=1+G_{\boldsymbol{k}}(\tau=0^{+})\label{eq:occ_Gf}
\end{equation}

Lastly, $\delta t$ (in Eq. (\ref{eq:vareps_def})) is also known,
in the Fock approximation, as a function of $G_{ij}(\tau=0^{+})$ (see Eq.~\eqref{eq:deltat}):
\begin{equation}
\delta t=-VG_{\langle ij\rangle}(\tau=0^{+})\label{eq:deltat_Gf}
\end{equation}

Putting (\ref{eq:occupancy_final}-\ref{eq:occ_Gf}-\ref{eq:deltat_Gf})
together and Fourier transforming, one gets

\begin{equation}
G_{\langle ij\rangle}(\tau=0^{+})=\frac{1}{2}\left(-\frac{t-VG_{\langle ij\rangle}(\tau=0^{+})}{U}\right).
\end{equation}
Solving for $G_{\langle ij\rangle}(\tau=0^{+})$, one gets:

\begin{equation}
G_{\langle ij\rangle}(\tau=0^{+})=-\frac{t}{2U-V}\label{eq:G_ij_final}
\end{equation}

and 
\begin{equation}
\delta t=\frac{tV}{2U-V}.\label{eq:deltat_final}
\end{equation}

This expression is used in Section~\ref{sec:insidethemottphase}.

\bibliographystyle{apsrev4-1}
\bibliography{refs}
\end{document}